\documentclass[superscriptaddress,prx,twocolumn]{revtex4-2}
\usepackage{graphicx}
\usepackage{amsmath,amssymb,physics,dsfont}
\usepackage{color}
\usepackage{hyperref}
\usepackage{microtype}
\usepackage{braket}
\usepackage{defs-private}
\newcommand{\revision}[1]{\color{black}#1 \color{black}}
\begin{document}

\title{Gutzwiller Hybrid Quantum-Classical Computing Approach for Correlated Materials}
\author{Yongxin Yao}
\email{ykent@iastate.edu}
\affiliation{Ames Laboratory, U.S. Department of Energy, Ames, Iowa 50011, USA}
\affiliation{Department of Physics and Astronomy, Iowa State University, Ames, Iowa 50011, USA}

\author{Feng Zhang}
\affiliation{Ames Laboratory, U.S. Department of Energy, Ames, Iowa 50011, USA}

\author{Cai-Zhuang Wang}
\affiliation{Ames Laboratory, U.S. Department of Energy, Ames, Iowa 50011, USA}
\affiliation{Department of Physics and Astronomy, Iowa State University, Ames, Iowa 50011, USA}

\author{Kai-Ming Ho}
\affiliation{Ames Laboratory, U.S. Department of Energy, Ames, Iowa 50011, USA}
\affiliation{Department of Physics and Astronomy, Iowa State University, Ames, Iowa 50011, USA}

\author{Peter P. Orth}
\email{porth@iastate.edu}
\affiliation{Ames Laboratory, U.S. Department of Energy, Ames, Iowa 50011, USA}
\affiliation{Department of Physics and Astronomy, Iowa State University, Ames, Iowa 50011, USA}

\begin{abstract}
	  Rapid progress in noisy intermediate-scale quantum (NISQ) computing technology has led to the development of novel resource-efficient hybrid quantum-classical algorithms, such as the variational quantum eigensolver (VQE), that can address open challenges in quantum chemistry, physics and material science. Proof-of-principle quantum chemistry simulations for small molecules have been demonstrated on NISQ devices. While several approaches have been theoretically proposed for correlated materials, NISQ simulations of interacting periodic models on current quantum devices have not yet been demonstrated. Here, we develop a hybrid quantum-classical simulation framework for correlated electron systems based on the Gutzwiller variational embedding approach. We implement this framework on Rigetti quantum processing units (QPUs) and apply it to the periodic Anderson model, which describes a correlated heavy electron band hybridizing with non-interacting conduction electrons. Our simulation results quantitatively reproduce the known ground state quantum phase diagram including metallic, Kondo and Mott insulating phases. This is the first fully self-consistent hybrid quantum-classical simulation of an infinite correlated lattice model executed on QPUs, demonstrating that the Gutzwiller hybrid quantum-classical embedding framework is a powerful approach to simulate correlated materials on NISQ hardware. This benchmark study also puts forth a concrete pathway towards practical quantum advantage on NISQ devices.
\end{abstract}

\maketitle

\section{Introduction}
Quantum computing holds the promise to revolutionize modern high-performance computations in physics by providing exponential speedups compared to currently known classical algorithms for a variety of important problems such as simulating interacting quantum models~\cite{feynman82qc,lloyd1996, Laflamme_QuantumFermionicSimulation, Laflamme_SimulatingPhysicsbyQuantumNetwork}. Accurately predicting the properties of competing phases or simulating the dynamics of interacting quantum mechanical many-body systems directly addresses grand challenges in quantum chemistry and materials science~\cite{dagotto2005complexity,RMPNonequilibriumDMFT,ultrafast_approach, quantuMaterialOnDemand,doe_qis}.

While not being fully fault-tolerant, the currently available noisy intermediate-scale quantum (NISQ) hardware~\cite{nisq} is still extremely powerful as recently demonstrated by the Google team~\cite{quantum_supremacy19}. As the number of coherent gate operations is limited, however, the development of resource efficient algorithms with sufficiently short quantum circuits is crucial in order to be able to tackle open scientific problems on NISQ devices. One example is the variational quantum eigensolver (VQE) algorithm to solve the eigenvalue problem~\cite{vqe, vqe_theory}. It has been successfully implemented on NISQ technology to compute the ground state energy of small molecules such as H$_2$, HHe$^+$, LiH and BeH$_2$ ~\cite{vqe, hardware_efficient_vqe, vqe_pea_h2, vqe_NaH}. The VQE algorithm adopts a hybrid quantum-classical approach which combines a quantum computation of a suitable cost function, such as the Hamiltonian, with a classical method for optimization. Instead of adiabatic state preparation followed by quantum phase estimation~\cite{asp,asp_ipea}, which requires deep circuits, VQE employs shallow variational circuits to evolve a chosen initial state into the target state. The cost function is then measured as a weighted sum of expectation values for associated Pauli terms. The variational parameters are classically optimized to minimize the cost~\cite{vqe, vqe_theory}.

Different forms of the variational circuit in VQE, e.g., the unitary coupled cluster ansatz (UCC)~\cite{vqe_theory,grimsleyAdaptiveVariationalAlgorithm2019,alan_ucc2018}, the qubit coupled cluster ansatz~\cite{qcc_scott2018,iqcc_scott2020}, or a trotterized adiabatic preparation ansatz~\cite{wecker2015_trotterizedsp}, have been discussed in the literature. A common issue of these variational ans\"atze is that the number of variational parameters rapidly increases with the number of orbitals in the calculation, which makes the generally non-convex classical optimization problem increasingly difficult to solve. This is further complicated by the presence of noise on real NISQ devices. While VQE algorithms have been demonstrated on NISQ devices for small molecules, computing properties of infinite periodic quantum materials requires further algorithmic development.

Various quantum algorithms for efficiently solving periodic correlated materials problems have been actively discussed in the literature. For example, it has been shown that using an adiabatic \revision{quantum variational} approach with a dual plane wave basis set leads to favorable scaling with respect to the number of basis orbitals for the circuit depth and the number of qubits required for periodic systems~\cite{qa_dualbasis}. The Jellium model has been proposed as a benchmark case for this approach on near-term devices. Another route is to follow the long tradition in correlated materials theory to map infinite periodic systems onto effective impurity models. Such an approach has been very successful in classical computing of correlated materials, e.g., the state-of-the-art dynamical mean-field theory (DMFT)~\cite{dmft_vollhardt,dmft_georges96,dmft_kotliar06,dmft_held07}. Quantum algorithms based on adiabatic state preparation and phase estimation have been suggested that solve for the impurity Green function repeatedly, upon reaching the convergence with the local lattice Green function~\cite{hybrd_dmft}. A hybrid quantum-classical approach based on a simplified two-site version of DMFT has also recently been described in Ref.~\cite{2sitedmft, hybrd_2sitedmft}, and it was proposed to use a generalized VQE method to find both ground and excited states of the impurity model~\cite{hybrd_2sitedmft2}. Although quite appealing, none of these proposed algorithms have yet been demonstrated on a real NISQ device, because the resources they require for the study of infinite periodic systems are still beyond the current technology~\cite{dmft_qcmin_jaderberg2020}.

In this paper we develop and demonstrate a novel resource-efficient hybrid quantum-classical algorithm that can simulate correlated materials on present-day NISQ devices. The algorithm is based on the Gutzwiller variational wavefunction for the interacting many-body ground state~\cite{ga, ga_bunemann1998} and thus captures correlations beyond a simple mean-field ansatz such as Hartree-Fock. However, it requires significantly less resources than DMFT and can thus be executed on current hardware. We have implemented this Gutzwiller quantum-classical embedding (GQCE) simulation framework on Rigetti's quantum cloud service (QCS) using PyQuil~\cite{pyquil_0,pyquil_1}, and used it to perform the first self-consistent calculations of an infinite periodic correlated electron model on a quantum computer. As a non-trivial benchmark study we investigate the periodic Anderson model (PAM) on Rigetti's Aspen-4 quantum device. Our results show that GQCE correctly describes the PAM ground state phase diagram, which contains Kondo insulator, correlated metal, and Mott insulator phases~\cite{dmft_vollhardt,pam_metal_insu,pam_logan2016}. In contrast to Hartree-Fock theory, the critical parameters for the associated quantum phase transitions are also accurately determined using GQCE. Our work demonstrates the current capabilities of NISQ devices in the simulation of correlated materials.

The GQCE approach is based on the powerful Gutzwiller variational embedding theory~\cite{ga_pu,ga_uo2}, which is known to be equivalent to the rotationally invariant slave-boson method in the saddle-point approximation~\cite{sb_kotliar86, bunemann_gasb, Lechermann_risb, ga_uo2}. The Gutzwiller embedding theory can capture many phenomena associated with strong local electron correlations such as  Mott-Hubbard transitions~\cite{BrinkmanRice, GutzwillerMottTosatti,ga_uo2}, unconventional superconductivity~\cite{GutzwillerSCBunemann, GutzwillerSCFabrizio, TDGutzwillerSC}, quantum spin liquids~\cite{gutzwillerRVB, GutzwillerSpinLiquid}, and topological phases~\cite{ga_smb6, GutzwillerWeyl, GutzwillerAFMChern}. When combined with \emph{ab initio} density-functional theory (DFT), the Gutzwiller approach is well suited for studying ground state properties of real correlated materials  ~\cite{ga_ce,ga_fe,ga_feas_bunemann,ga_pu,ga_uo2,ga_smb6,ga_tmo,comsuite}.

Similar to DMFT, the Gutzwiller embedding method maps the infinite interacting lattice model onto an effective impurity problem consisting of a cluster of correlated orbitals embedded in a self-consistent medium. Unlike DMFT, however, which solves for the fully frequency dependent impurity self-energy, the Gutzwiller theory requires only the ground state single-particle density matrix of the embedding correlated cluster. In practice, the Gutzwiller embedding approach amounts to finding a self-consistent solution of a set of coupled eigenvalue equations. The method is therefore ideally suited to be formulated as a hybrid quantum-classical algorithm, where the ground state of the correlated impurity cluster can be efficiently determined using VQE. 

The GQCE calculations share the favorable polynomial system size scaling of VQE in solving the interacting embedding Hamiltonian. Therefore, GQCE promises to be able to consider larger embedding clusters, which take multi-orbital or spatial correlations into account. This is necessary to describe the non-local electronic order parameters such a $d$-wave superconductivity~\cite{cluster_dmft_afm_dwave_lichtenstein2000, cluster_dmft_d_wave_Keller00, cluster_dmft_d_wave_Emanuel13}, the impact of short-range fluctuations on electronic properties~\cite{cluster_dmft_kotliar08}, and composite order parameters of vestigial phases~\cite{vestigial_order_fernandes19}. In the near term, a robust VQE solution of a 28-qubit Hubbard-type Hamiltonian, which is equivalent to a Gutzwiller embedding Hamiltonian of a single $f$-orbital site in rare-earth and actinide materials, would bring the capabilities of GQCE calculations on NISQ devices to the verge of what is currently possible on classical computers, thus demonstrating practical quantum advantage.

\section{Hybrid Gutzwiller embedding framework}
In this section, we introduce the key components of the hybrid quantum-classical Gutzwiller embedding framework and describe its implementation on NISQ QPUs. We highlight several advantages of the quantum algorithm compared to its purely classical counterpart, in particular the favorable polynomial, compared to exponential, scaling of the algorithmic complexity with the size of the real-space embedding cluster. This is important as the Gutzwiller embedding method systematically approaches the exact solution as the cluster size increases.  Larger clusters also allow to describe qualitatively new physical phenomena, for example, spatially extended order parameters and correlations.

\subsection{General GQCE framework}
\revision{
The GQCE framework is based on the Gutzwiller quantum embedding theory to calculate ground state properties of correlated electron materials~\cite{ga_pu,ga_uo2}, which was shown to be equivalent to the rotationally invariant slave-boson (RISB) method in the saddle-point approximation~\cite{sb_kotliar86, bunemann_gasb, Lechermann_risb, ga_uo2}. As the formalism of Gutzwiller-slave-boson approach has been extensively presented previously~\cite{ga_pu,ga_uo2,sb_kotliar86,bunemann_gasb, Lechermann_risb}, we here focus on the novel hybrid quantum-classical implementation of the method.

Consider a generic multi-band Hubbard Hamiltonian with local onsite screened Coulomb interactions for periodic systems
\bea
\h &=& \sum_{\bk}\sum_{\mu\nu} t_{\bk\mu\nu}\, \cc_{\bk\mu}\ca_{\bk \nu} \notag \\ &+&\frac{1}{2}\sum_{\bR l}\sum_{p q p' q'} \sum_{\s\s'}V_{p q p' q'}^l \cc_{l p \s} \cc_{l p' \s'} \ca_{l q' \s'} \ca_{l q \s}, \label{eq: hamgen0}
\eea
where $\bk$ is the crystal momentum, conjugate to the unit cell position vector $\bR$. The unit cell can refer to a primitive unit cell or a supercell. $\mu,\nu$ are composite indices of the lattice basis site and orbital, with spin included unless explicitly labelled by $\s$. Orbitals include both uncorrelated orbitals with negligible screened Coulomb interactions such as $s$ and $p$-orbitals, and correlated orbitals  $\{\phi_{l p} \}$ with significant Coulomb interactions such as $d$ and $f$-orbitals, which are explicitly labelled by $p$ and $q$ at the $l^\text{th}$ correlated site, with the screened Coulomb integral expressed as
\be
V_{p q p' q'}^l=\iint d\br d\br'\phi_{l p}^*(\br) \phi_{l q}(\br) V(\abs{\br - \br'})\phi_{l p'}^*(\br') \phi_{l q'}(\br'). \label{eq: Coulomb integral}
\ee
All the one-body terms, such as hopping, crystal-field splitting and spin-orbit coupling, are included in $t_{\bk\mu\nu}$. Note that the Hamiltonian in Eq.~\eqref{eq: hamgen0} can describe idealized lattice models, such as the Hubbard model, but also real (multi-orbital) materials with parameters obtained from mean-field electronic structure calculations such as density functional theory (DFT).

To facilitate later discussions, we recast the Hamiltonian~\eqref{eq: hamgen0} in the following form:
\be
\h = \sum_{\bk}\sum_{\mu\nu} \epsilon_{\bk\mu\nu}\, \cc_{\bk\mu}\ca_{\bk \nu} +\sum_{\bR i}\h_i^\text{loc}[\bR]. \label{eq: hamgen}
\ee
Here we group the correlated orbital sites into clusters labelled by $i$, which can include relevant uncorrelated orbitals as well. For example, a cluster $i$ can contain a fractionally occupied $d$-shell at correlated site $l$, but may also include additional neighboring sites and orbitals to form a real-space multi-site cluster. The local interacting Hamiltonian $\h_i^\text{loc}[\bR]$ is defined at the $i^\text{th}$ cluster, which is identical at different unit cell positions $\bR$. It can generally be written as
\be
\h_i^\text{loc} = \sum_{\alpha\beta}t_{i\alpha\beta}\cc_{i\alpha}\ca_{i\beta} + \frac{1}{2}\sum_{\alpha\beta\gamma\delta}\sum_{\s\s'}V_{\alpha\beta\gamma\delta}^i \cc_{i\alpha\s} \cc_{i\gamma\s'} \ca_{i\delta\s'} \ca_{i\beta\s}.
\ee
The local Hamiltonian includes all the associated one-body and two-body terms. $\alpha, \beta, \gamma, \delta$ label the orbital sites in the correlated cluster, with spin included unless explicitly labelled. The Coulomb matrix element $V_{\alpha\beta\gamma\delta}^i$ is nonzero only  within the same correlated shell at site $l$ present in the $i^\text{th}$ cluster, as described by the nonzero elements $V_{p q r s }^l$ (Eq.~\eqref{eq: Coulomb integral}). Accordingly, $\epsilon_{\bk\mu\nu}$ is equal to $t_{\bk\mu\nu}$ in Eq.~\eqref{eq: hamgen0} subtracting the one-body components defined on correlated clusters, which have been merged to the local Hamiltonians in the second term of Eq.~\eqref{eq: hamgen}. 

The Gutzwiller variational wavefunction (GWF) is employed to evaluate the ground state property of the Hamiltonian~\eqref{eq: hamgen}, which takes the form:
\be
\ket{\Psi_\text{G}} = \prod_{\bR i} \hat{\P}_{\bR i} \ket{\Psi_0}, \label{eq: gwf}
\ee
with a noninteracting wavefunction $\ket{\Psi_0}$ and correlation projector
\be
\hat{\P}_{\bR i} = \sum_{A,B}[\Lambda_i]_{A B}\Pj{A,\bR i}{B,\bR i}. \label{proj}
\ee
The labels $A,B$ enumerate the complete Fock states of the local Hilbert space defined by filling the $N_i^\text{so}$ spin-orbitals in the $i^\text{th}$-cluster with number of electrons $N_i^\text{e} \in [0, N_i^\text{so}]$. The variational parameter matrices $\{ \Lambda_i \}$ are introduced in the Gutzwiller projector $\hat{\P}_{\bR i}$ to optimize local correlated sectors of the noninteracting $\ket{\Psi_0}$. 
}

\begin{figure}[ht]
	\centering
	\includegraphics[width=\linewidth]{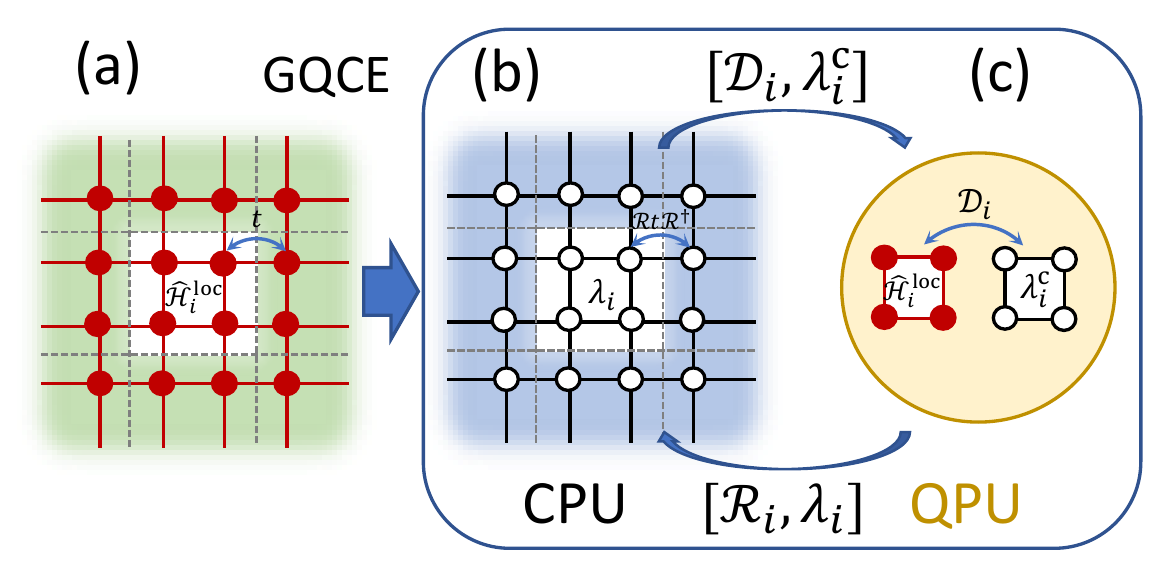}
	\caption{
	\textbf{Schematic illustration of the generic GQCE framework.} Panel (a) shows an interacting quantum lattice model, exemplified by a real-space cluster $i$ described by $\h_i^\text{\text{loc}}$ and hopping amplitude $t$ to other sites. This interacting lattice model is self-consistently mapped to a noninteracting quasiparticle lattice model shown in panel (b) and a finite-size, interacting embedding model (representing the $i$-cluster) that is coupled to a noninteracting bath of the same size [see panel (c)]. The GQCE method requires finding a self-consistent solution of the ground state of the coupled quasiparticle and embedding Hamiltonians. Within GQCE, the interacting embedding Hamiltonian is solved on QPUs using quantum algorithms such as VQE. The quasiparticle Hamiltonian can be efficiently simulated on classical processing units (CPUs).
	}
	\label{figure0}
\end{figure}

As illustrated in Fig.~\ref{figure0}, within the Gutzwiller embedding theory, minimizing the total energy with respect to the GWF, $E_G = \min_{\{\ket{\Psi_0},\Lambda_i\}} \braket{\Psi_G | \hat{\mathcal{H}} | \Psi_G}$, leads  to a set of coupled eigenvalue equations at the Gutzwiller-rotationally invariant slave-boson (GRISB) level~\cite{ga_pu, ga_uo2}:
\be
\h_\text{G}^\text{qp}[\R,\R^\dagger; \lambda]\ket{\Psi_0} = E^\text{p} \ket{\Psi_0}, \label{eq: qpe}
\ee
which describes a noninteracting quasiparticle system, and 
\be
\h^\text{emb}_i [\D,\D^\dagger; \lambda^\text{c}]\ket{\Phi_i} = E_i^\text{c} \ket{\Phi_i}, \label{eq: embe}
\ee
which describes the interacting embedding electron system of the $i^\text{th}$-cluster. 

More specifically, the quasiparticle Hamiltonian takes a quadratic form:
\be
\h_\text{G}^\text{qp}[\R,\R^\dagger; \lambda] \equiv \hat{\T_\text{G}}[\R,\R^\dagger] 
    + \sum_{i a b}[\lambda_i]_{ab}\fc_{i a}\fa_{i b}, \label{eq: qph}
\ee
with the renormalized kinetic energy term defined as
\be
\hat{\T_\text{G}}[\R,\R^\dagger] = \frac{1}{N_\bk}\sum_{\bk}\sum_{\mu\nu}\sum_{a b} \epsilon_{\bk\mu\nu} \R_{a\mu}\R_{\nu b}^\dagger \fc_{\bk a}\fa_{\bk b},
\ee
by a simple rule $\ca_{\bk\mu} \rightarrow \sum_a \R^\dag_{\mu a} \fa_{\bk a}$. Here $N_\bk$ is the total number of $\bk$-points, and the square-matrix $\R$ is the so-called Gutzwiller renormalization factor for the noninteracting quasiparticles represented by $\fa$operators labelled by indices $a$ and $b$, which run through the same number of spin-orbitals as the interacting labels $\alpha, \beta$ on the $i^\text{th}$-cluster. The Gutzwiller quasi-particle spectral weight is given by $Z\equiv \R^\dagger \R$, which is a measure of the electron correlation effect and characterizes the Mott transition by some vanishing components~\cite{ga_uo2, BrinkmanRice}. The gravity center of the quasiparticles is further renormalized by a matrix $\lambda$. The embedding Hamiltonian~\eqref{eq: embe} is given by
\bea
\h^\text{emb}_i [\D,\D^\dagger; \lambda^\text{c}] &=& \h^\text{loc}_i 
    + \sum_{a \alpha}\left( [\D_i]_{a \alpha} \cc_{i \alpha} \fa_{i a} + h.c.\right)
    \notag \\
    & & +\sum_{a b}[\lambda_i^\text{c}]_{a b} \fa_{i b} \fc_{i a}\,.
    \label{eq: hembg}
\eea
It describes an interacting subsystem, namely the $i^\text{cluster}$ with Hamiltonian $\h^\text{loc}_i$, coupled to a finite, noninteracting bath, which is characterized by the matrix $\lambda^\text{c}_i$. The hybridization coupling strength is given by the matrix $\D_i$. Here $\alpha$ labels the spin and orbitals in the $i^\text{th}$-cluster, and $a,b$ are the spin-orbital labels of the bath sites. 

\subsection{Essentials of the GQCE algorithm}
The GQCE algorithm is beyond the conventional mean-field theory such as Hartree-Fock, whose solution is completely determined by an effective single-particle Hamiltonian. The GQCE calculation amounts to self-consistently solving a set of eigenvalue equations~\eqref{eq: qpe} and \eqref{eq: embe}, which describe an interacting electron subsystems embedded in a noninteracting quasiparticle bath. 

The iterative procedure starts with finding the ground state wavefunction $\ket{\Psi_0}$ of the quasi-particle Hamiltonian $H_\text{G}^\text{qp}$ \eqref{eq: qph} defined by an initial guess of $\{ \R, \lambda\}$. The noninteracting wavefunction $\Psi_0$ determines the matrices $\{\D, \lambda^\text{c} \}$ entering the embedding Hamiltonians $\{\h_i^\text{emb}\}$ \eqref{eq: hembg}, which is subsequently solved for the ground state wavefunctions $\{\Phi_i\}$. To determine whether self-consistency is reached, one calculates the expectation value $E_i^\text{c}$ and the single-particle density matrix for each symmetrically-inequivalent embedding Hamiltonian $\h^\text{emb}_i$. Comparison to the corresponding quantities of the quasiparticle Hamiltonian allows to define a vector error function that vanishes at the self-consistent solution of two coupled eigenvalue problems: 
\bea
\left[ \F_i^1\right]_{a\alpha} &\equiv& \sum_c\left[\Delta_{\text{p} i}(1-\Delta_{\text{p} i})\right]^{-\frac{1}{2}}_{c a}  \Av{\Phi_i}{\cc_{i\alpha}\fa_{i c}} - [\R_{i}]_{a\alpha}, \notag \\
\left[ \F_i^2\right]_{ab} &\equiv& \Av{\Phi_i}{\fa_{i b}\fc_{i a}} - [\Delta_{\text{p}i}]_{ab}, \label{err}
\eea
where $[\Delta_{\text{p} i}]_{a b}=\Av{\Psi_0}{\fc_{i a}\fa_{i b}}$ is the quasiparticle density matrix. Various numerical methods can be used to solve this set of nonlinear equations given the above vector error function~\cite{hybr, 2020SciPy-NMeth}, and more details are given below.

In the above iterative procedure, the ground state solution of the noninteracting quasiparticle Hamiltonian $H_\text{G}^\text{qp}$ can be efficiently calculated on classical computers, \revision{with computational time scaling as $\bigO(N^3)$ with respect to the number of orbitals $N$ in the unit cell.} The embedding Hamiltonian $\h_i^\text{emb}$ on the other hand describes an interacting finite size system. Therefore, exact diagonalization (ED) is used to find its ground state. The classical computational resources and time required to determine the ground state using ED scale exponentially with the single-particle basis dimension $N_i^\text{emb}$ of the embedding Hamiltonian~\eqref{eq: hembg}. In practice, a general embedding Hamiltonian of an $f$-electron system, which is represented by 14 ($f$-shell, $\ca_{i\alpha}$) + 14 (bath, $\fa_{i a}$) = 28 spin-orbitals, is close to the limit that classical computers can handle~\cite{ga_pu, ga_uo2}.

Importantly, the computational accuracy of the embedding method can be systematically improved by increasing the size of the Gutzwiller projector $\P_i$ to act on a larger correlated cluster~\cite{grg, gga_lanata2017, ga_dmet}, which enlarges the orbital dimension of the embedding Hamiltonian. The exponential scaling of the ED solver with the orbital dimension therefore imposes a limit to the maximal accuracy that the Gutzwiller embedding approach can achieve on classical computers. To overcome this fundamental limitation, we propose to efficiently solve $\h_i^\text{emb}$ on QPUs using quantum algorithms such as VQE. This scheme makes full use of the advantageous linear scaling of the required number of qubits when increasing the size of the embedding Hamiltonian. Therefore, simulations using 20 (28) qubits can fully capture the complete manifold of $\h_i^\text{emb}$ of local $d$-orbitals ($f$-orbitals). 

\revision{More specifically, in this work we use VQE with UCC ansatz at single and double excitation level (UCCSD)~\cite{vqe_theory, vqe_pea_h2, alan_ucc2018} to solve for the ground state energy and one-particle density matrix (OPDM) of the embedding Hamiltonian~\eqref{eq: hembg}. The VQE-UCCSD calculation typically starts with a Hartree-Fock (HF) calculation, and transforms $\h_i^\text{emb}$~\eqref{eq: hembg} from atomic orbital basis to molecular orbital ($\phi$) representation, which can be cast in the form of a conventional molecular Hamiltonian (apart from a constant):
\be
\h^\text{emb} = \sum_{r s}h^\text{(1)}_{r s}\hat{\phi}_{r}^\dag \hat{\phi}_{s} + \frac{1}{2}\sum_{r s r' s'}\sum_{\s \s'} h^\text{(2)}_{r s r' s'}\hat{\phi}_{r\s}^\dag \hat{\phi}_{r'\s'}^\dag \hat{\phi}_{s'\s'} \hat{\phi}_{s\s}. \label{eq: hembg_hf}
\ee
Here, the cluster index $i$ is omitted for simplicity. $r$ and $s$ label HF molecular orbitals, with spin included unless explicitly labelled. The ladder operator $\hat{\phi}_r$ is a linear combination of $\{\ca_{i\alpha}, \fa_{i a} \}$ due to the basis transformation. The scalability of VQE-UCCSD has been extensively discussed in the literature, for example, in Ref.~\cite{alan_ucc2018}. For the above embedding Hamiltonian with $N_i^\text{emb}$ spin-orbital sites and fixing the number of electrons to half-filling, the number of gates scales as $\bigO[(N_i^\text{emb})^5]$ using Jordan–Wigner transformation~\cite{map_jw}. VQE-UCCSD with a Bravyi-Kitaev mapping is expected to have similar circuit complexity due to the implementation of Pauli rotation gates in the exponential form~\cite{map_bk, alan_ucc2018}. The ground state energy is obtained through Hamiltonian averaging~\cite{vqe_theory}. Recently, it has been shown that the number of partitions for distinct measurement circuits can be reduced to be $\bigO(N_i^\text{emb})$ with the gate counts to $(N_i^\text{emb})^2/4$ by employing the low-rank tensor factorization of the Hamiltonian coefficients $h^\text{(1)}$ and $h^\text{(2)}$,~\cite{tensordecomposition_u, tensordecomposition_qs, hamiltonianfactorization}. Importantly, only one-qubit $Z$ and two-qubit $Z Z$ operators need to be measured as a result of tensor-factorization, hence the exponential growth of measurement error with respect to the Pauli operator length is reduced to minimum~\cite{hamiltonianfactorization}. Furthermore, it has been shown that the number of repeated measurements to reach an accuracy of $\varepsilon$ of the total energy is much reduced from the upper bound according to the Hamiltonian coefficients, $\left( \sum_n \abs{\omega_n}/\varepsilon \right)^2$, for Hamiltonian $H=\omega_n P_n$ as a sum of Pauli terms $\{P_n \}$ with coefficients $\{\omega_n \}$~\cite{hamiltonianfactorization, wecker2015_trotterizedsp}.  Compared with typical ground state energy calculations using VQE-UCCSD in quantum chemistry, the expectation values of OPDM operators $\{\hat{\phi}_{r}^\dag \hat{\phi}_{s} \}$ will only be measured with the final optimized VQE ansatz. Following the above matrix factorization of $h^\text{(1)}$ for the sets of commuting OPDM operators, the number of additional measurement circuits for OPDM also scales as $\bigO(N_i^\text{emb})$, along with the favorite scaling of the number of measurements to achieve certain accuracy.
}
\label{sec: scaling}

We emphasize that the Gutzwiller embedding theory only requires finding the ground state energy and OPDM, which can be implemented successfully on present-day QPUs as we show below. \revision{In contrast, DMFT often has a more complicated embedding Hamiltonian. In the simple version of two-site DMFT~\cite{2sitedmft, hybrd_2sitedmft}, the embedding Hamiltonian is of the same complexity as that of GQCE. However, DMFT requires determining the full frequency dependent embedding Green's function, which is challenging on current NISQ hardware, since it requires simulating excited states as well~\cite{dmft_qcmin_jaderberg2020}.}

\revision{Let us briefly comment on the possibility of using VQE to directly optimize the GWF without resorting to the GRISB (saddle-point) approximation~\cite{sb_kotliar86, bunemann_gasb, Lechermann_risb, ga_uo2}. The total energy $\langle \Psi_\text{G} | \h |\Psi_{\text{G}}\rangle$ can in principle be directly evaluated by Hamiltonian averaging by preparing the GWF on QPUs without resorting to the GRISB approximation used in GQCE. The GWF state could then be optimized subsequently using an algorithm such as VQE. However, this GWF-VQE approach requires a large number of qubits, equal to the number of spin-orbital sites in the large Born–von Karman supercell of a periodic system~\cite{ashcroft1976solid}. In contrast, within GQCE (which exploits the GRISB) one only needs to find the ground state of the much smaller many-body embedding model defined in Eq.~\eqref{hemb}. In addition, the variational degrees of freedom of the GWF, $\{\Lambda_i \}$, represent a high-dimensional parameter space, such that classical optimization poses a serious challenge. In contrast, within GQCE, the complex optimization problem is mapped to a self-consistent solution of two coupled ground state eigenvalue problems, which is numerically much more straightforward. Although the total energy functionals of GQCE and GWF-VQE generally differ at finite dimension $d$, they were shown to become identical in the infinite dimension limit $d \rightarrow \infty$~\cite{ga_infdvollhardt}. The benchmark calculation of the periodic Anderson model that we present below is performed in the $d \rightarrow \infty$ limit. In addition, it was shown that even for finite dimensional systems, GRISB often maintains the variational nature in practice by producing an upper bound of the energy that converges to the exact answer with increasing size of the Gutzwiller projector~\cite{ga_dmet}. To conclude, the GQCE approach is more NISQ friendly than GWF-VQE and will thus be pursued in the following. 
}
\label{variational_in_practice}

\subsection{Implementation of GQCE simulation framework}
The GQCE framework is built on the open-source CyGutz package, which is an implementation of the Gutzwiller embedding approach in classical computers~\cite{cygutz,comsuite}. Here we have developed the quantum computing module of GQCE using both IBM Qiskit and Rigetti's Forest SDK~\cite{Qiskit, pyquil_0, pyquil_1}, which is released as an open-source code~\cite{pygqce}. The statevector simulator in IBM Qiskit and the wavefunction simulator in Forest SDK have been employed for noiseless simulations. The GQCE calculations on real quantum devices are conveniently performed through the quantum cloud service (QCS) by Rigetti. The QCS provides a quantum machine image that is co-located with the quantum infrastructure, which allows fast virtual execution of hybrid quantum-classical programs at low latency cost. Platform-level optimizations of parametric compilation and active qubit reset, which dramatically reduce the latency in the QCS platform, have been utilized in our GQCE calculations. We employ the readout symmetrization and error mitigation techniques for the measurements, as implemented in reference~\cite{pyquil_1}, \revision{where error rates are first characterized for the symmetrized readout, and the measured observable expectation values are rescaled accordingly. Although this readout error mitigation is not scalable due to the exponential growth of the number of measurement circuits with the size of the Pauli term for calibrations, all the necessary observable measurements in GQCE can be reduced to one or two qubits by adopting the low-rank factorization technique as discussed in Sec.~\ref{sec: scaling}.} Additional error mitigation approaches, such as Richardson extrapolation techniques, have been proposed and experimentally realized recently~\cite{kandala_error_qc_2019, larose2020mitiq}. As demonstrated below, use of these more advanced strategies is not necessary for the benchmark calculations with single-site decoupling here, but will be advantageous for larger embedding clusters.

To determine convergence of the self-consistency loop, we monitor the error vector function $\F$~\eqref{err}, which describes the change of the trial solution after one iteration of the self-consistency loop. If $\F$ can be evaluated accurately, the modified Powell hybrid method~\cite{hybr} can be the method of choice to find a self-consistent solution of the coupled eigenvalue problem, as practiced in references~\cite{ga_ce, ga_pu, ga_uo2, ga_tmo}. The Powell method employs information about the numerical Jacobian. Since the noise level of current quantum devices due to gate infidelities and decoherence is significant~\cite{kandala_error_qc_2019}, $\F$ cannot be accurately calculated on noisy QPUs. \revision{In practice, the ``exciting-mixing'' method
performs sufficiently well for our purposes and we use it to solve the root problem of the noisy nonlinear equations. It replaces numerical evaluations of the Jacobian by a self-tuned diagonal Jacobian approximation, and it implemented in the SciPy library~\cite{2020SciPy-NMeth}, } We demonstrate that VQE calculations performed on Rigetti's Aspen-4 QPU with standard readout symmetrization and calibration yields sufficiently accurate results to reach self-consistency of the GQCE calculation.

\section{GQCE solution of the periodic Anderson model}
In this section, we present fully self-consistent GQCE calculations of the infinite PAM on Rigetti's Aspen-4 quantum 
processing unit (QPU). This demonstrate the feasibility of the GQCE framework on present-day NISQ hardware. Here, we focus on the single-site embedding version of the GQCE method as our goal is to carefully benchmark this new framework. In the future, larger multi-site and multi-orbital embedding Hamiltonians and interfacing GQCE with density-functional theory will be able to address more realistic models of correlated materials.

\begin{figure}[t!]
	\centering
	\includegraphics[width=\linewidth]{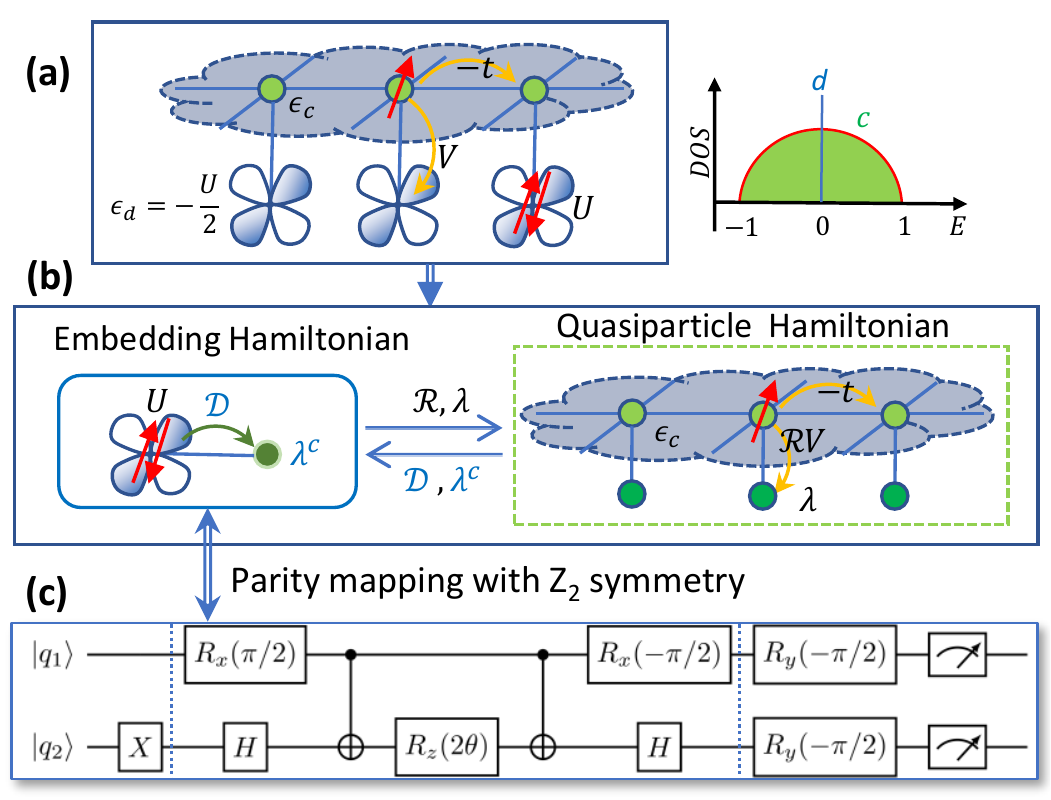}
	\caption{
	\textbf{Illustration of Gutzwiller hybrid quantum-classical embedding algorithm used in solving the periodic Anderson model (PAM).} (a) Sketch of the PAM on the Bethe lattice, together with the decoupled density of states of the itinerant $c$-band (semi-circle) and the correlated $d$-orbital ($\delta$-function). (b) Schematic view of the coupled eigenvalue problems. It involves an interacting quantum many-body embedding Hamiltonian, which is solved using VQE, and a non-interacting effective quasi-particle Hamiltonian, which results from the Gutzwiller variational ansatz. The model parameters are defined in the text. (c) A quantum circuit to solve for the ground state of the embedding Hamiltonian using a variational unitary coupled cluster (UCC) quantum eigensolver. The circuit includes three parts: initial HF state preparation, UCC ansatz, and a measurement of Pauli term $X_0 X_1$, as indicated by the vertical dotted lines.
	}
	\label{figure2}
\end{figure}

To perform a first non-trivial benchmark study of GQCE for infinite systems, we consider the periodic Anderson model (PAM) on the Bethe lattice in infinite dimension, as illustrated in Fig.~\ref{figure2}(a). The system is described by a Hamiltonian composed of an itinerant $c$-band, a local interacting $d$-orbital and onsite hybridization between them~\cite{pam_logan2016},
\be
\h=\h_c+\h_d+\h_\text{hyb},
\ee
where
\be
\h_c=\sum_{\bR\s}\epsilon_c\cc_{\bR\s}\ca_{\bR\s} -
    \sum_{\av{\bR,\bR'},\s}t\cc_{\bR\s}\ca_{\bR'\s},
\ee
\be
\h_\text{hyb}=\sum_{\bR\s}V\left( \dc_{\bR\s}\ca_{\bR\s} + h.c.\right).
\ee
and
\be
\h_d=\sum_{\bR}\h^\text{loc}[\bR],
\ee
with
\be
\h^\text{loc}[\bR]=\epsilon_d\dc_{\bR\s}\da_{\bR\s} +
    \frac{1}{2}\sum_{\s}U\dc_{\bR\sbar}\da_{\bR\sbar}\dc_{\bR\s}\da_{\bR\s}.
\ee
The center of the itinerant $c$-band is given by $\epsilon_c$ and the energy of the correlated $d$-orbital is given by $\epsilon_{d}$. $U$ denotes the intra-orbital Hubbard interaction parameter on the $d$-orbital, and $V$ the on-site hybridization strength between $c$- and $d$-electrons. The spin index $\sbar$ indicates the opposite of $\s$. As the $d$-orbital is the only correlated group in the unit cell labelled by $\bR$, the group index $i$ is skipped in the model. For reference, the Fourier transformation of $\h_c + \h_\text{hyb}$ to the momentum $\bk$-space constitutes the first part of the generic Hamiltonian~\eqref{eq: hamgen}. On a Bethe lattice in infinite dimensions or with infinite nearest-neighbor connectivity, the conduction band density of states (DOS) takes the semi-circular form $\rho_{c}(\epsilon)=\frac{2}{\pi D}\sqrt{1-(\epsilon/D)^2}$, where $D$ is the half band width. We set $D=1$ in the following calculations. The model hosts a diversity of paramagnetic electronic phases: a metal, band insulator, Kondo insulator and Mott insulator. The different phases are separated by quantum phase transitions. The model has been extensively studied in the literature~\cite{dmft_vollhardt,pam_metal_insu,pam_logan2016}, and highly accurate numerical results have been obtained using DMFT~\cite{dmft_vollhardt,dmft_georges96,dmft_kotliar06,dmft_held07}, which becomes exact for systems in infinite dimension. This makes the PAM model on the Bethe lattice an ideal benchmark model for hybrid quantum-classical calculations of infinite correlated electron systems on NISQ devices.

In this work, we choose the particle-hole symmetric point of $\h_{d}$ with Fermi level at $0$, \textit{i.e.}, we set $\epsilon_{d}=-U/2$, and also fix $V=0.4$ and $U=2$. We determine the ground state phase diagram as a function of conduction band energy center $\epsilon_c$. In this parameter space, the system starts with a Kondo insulating (KI) phase for $\epsilon_c=0$. With increasing $\epsilon_c$ it first transforms into a metallic (M) phase and finally enters the Mott-Hubbard insulating (MI) regime undergoing a metal-insulator transition~\cite{pam_logan2016}. From DMFT calculations using numerical renormalization group (NRG) as an impurity solver~\cite{nrg, pam_logan2016}, the zero temperature quantum phase transitions occur at the critical values of $\epsilon^{\text{KI-M}}_c = 0.07$ and $\epsilon^{\text{M-MI}}_c = 1.08$. These values can be considered as numerically exact for this model.

To study the PAM, we consider a GWF \eqref{eq:  gwf} with the correlation projector acting on the Hilbert space spanned by the single particle $d$-orbitals. This leads to a set of coupled eigenvalue equations governed by a Gutzwiller embedding Hamiltonian, which provides an accurate description of local electron correlations, together with a non-interacting effective quasi-particle Hamiltonian. The method is schematically illustrated in Fig.~\ref{figure2}(b).

The Gutzwiller embedding Hamiltonian \eqref{eq: hembg} holds a specific form, apart from a constant, as
\bea
\h^\text{emb}&=&\sum_{\s}\epsilon_{d}\dc_{\s}\da_{\s} +
    U\dc_{\up}\da_{\up}\dc_{\dw}\da_{\dw} \notag  \\
    && +\sum_{\s}\left(\D\dc_{\s}\fa_{\s}+h.c.\right)
    -\sum_{\s}\lambda^{c}\fc_{\s}\fa_{\s}\,.
    \label{hemb}
\eea
Here, $\D$ denotes the coupling strength between the $d-$orbital and a non-interacting bath orbital $f$ with energy level $-\lambda^{c}$. The general local one-body matrix such as the kinetic energy renormalization matrix $\R$ and the coupling matrix $\D$, have a $2\times2$ diagonal form with degenerate diagonal elements due to spin-rotation symmetry in the paramagnetic state.

The non-trivial task is to solve for the ground state of the interacting embedding Hamiltonian $\h^\text{emb}$ \eqref{hemb} using VQE on quantum devices. We first transform $\h^\text{emb}$ to a molecular orbital representation, using the orbitals obtained from a spin-restricted Hartree-Fock (HF) calculation. Then, the Hamiltonian is written in a qubit representation via standard parity mapping~\cite{map_bk, map_three}. Since the ground state at half-filling $N_e = 2$ is restricted to total spin $S=0$, the embedding Hamiltonian can be represented in a two-qubit basis exploiting $\mathbb{Z}_2$ symmetries as
\bea
\h^\text{emb} &=& g_0 \mathds{1} + g_1 (Z_0-Z_1) + g_2 (X_0+X_1) + g_3 Z_0 Z_1 \notag \\
&+& g_4 (X_0Z_1 - Z_0 X_1) + g_5 X_0 X_1\,.
    \label{qham}
\eea
Here $X_i,Y_i$ and $Z_i$ are Pauli operators acting on qubit $i$, and the parameters $\{ g_\alpha\}$ are determined by parameters of the embedding Hamiltonian~\eqref{hemb} and the form of the HF molecular orbitals. \revision{(see supplemental material for details.)} The asymmetric two-site embedding Hamiltonian is slightly more complex than that of the hydrogen dimer H$_2$, which is a widely used example for the application of VQE in quantum chemistry~\cite{vqe_pea_h2}.

\begin{figure*}[t!]
	\centering
	\includegraphics[width=.9\linewidth]{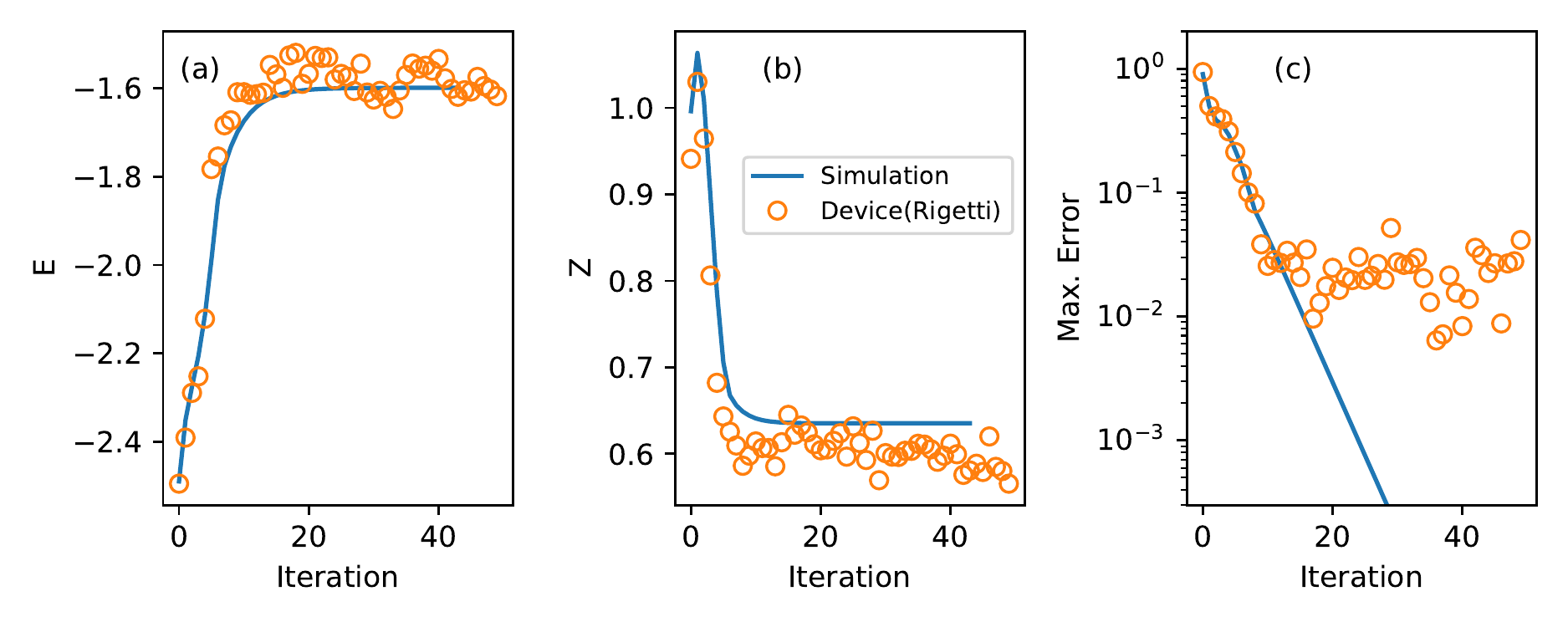}
	\caption{
	\textbf{Convergence behaviour of GQCE iterative calculations on Rigetti's Aspen-4 quantum device.} (a) Evolution of the system total energy, (b) kinetic energy renormalization $Z$-factor, and (c) the maximal element of the error vector $\F$ as a function of iteration number in solving the set of Gutzwiller nonlinear self-consistency equations. Results obtained from a calculation on Aspen-4 (orange circles) are compared to noiseless simulations using the statevector approach (blue line).
	}
	\label{figure3}
\end{figure*}

To find the ground state energy and single-particle density matrix, we use VQE with an unitary coupled cluster (UCC) ansatz~\cite{vqe}. For a two-electron system, the UCCSD ansatz at single and double excitation level is known to be exact. The single-excitation has no contribution to the ground state energy according to Brillouin theorem\cite{piela_ideas_qc_2020}. Importantly, the UCC ansatz can be reduced to a particularly simple form in two-qubit representation using a parity transformation~\cite{map_bk, map_three}
\bea
\ket{\Psi^{\text{ucc}}(\theta)} &=& e^{-i\theta Y_0 X_1}\ket{01},
\eea
where $\theta \in \left[-\pi, \pi\right]$ is a variational parameter and $\ket{01}$ is the spin-restricted HF ground state wave function. It is obtained by standard self-consistent calculations using a quantum chemistry PySCF package, which efficiently run on classical computers~\cite{pyscf_sun2018}.

To get the expectation value of the embedding Hamiltonian \eqref{qham} under the UCC wave function on quantum computers, we group Pauli terms that are diagonal in a common tensor-product basis. A typical quantum circuit, composed of the initial HF state preparation, UCC ansatz, and a measurement of Pauli term $X_0 X_1$, is shown in Fig.\ref{figure2}(c). In addition to Pauli terms contained in $\h^{\text{emb}}$, the Pauli term $Y_0$ is also measured with the optimized UCC ansatz to derive the OPDM of the embedding system. The VQE code is developed based on a quantum computing library pyQuil~\cite{pyquil_0, pyquil_1}, where we use a simultaneous perturbation stochastic approximation algorithm to optimize the noisy objective function on real quantum computing devices~\cite{spsa}.

The quantum processing unit (QPU) used in this study is Aspen-4. The device contains 13 qubits in total, among which we choose qubit 0 and 1 for the calculations. The associated two-qubit CZ-gate, which is one controlling factor for the noise level of the calculation results, has a fidelity of about 95\%. 

\section{Quantum computing results of periodic Anderson model}
The GQCE calculations on the PAM model are carried out in two ways. First, we use a statevector simulator, which represents an ideal fault-tolerant quantum computer with an infinite number of measurements. Second, we use two qubits on Rigetti's Aspen-4 quantum device, which contains 13 qubits in total.
\begin{figure*}[t!]
	\centering
	\includegraphics[width=.9\linewidth]{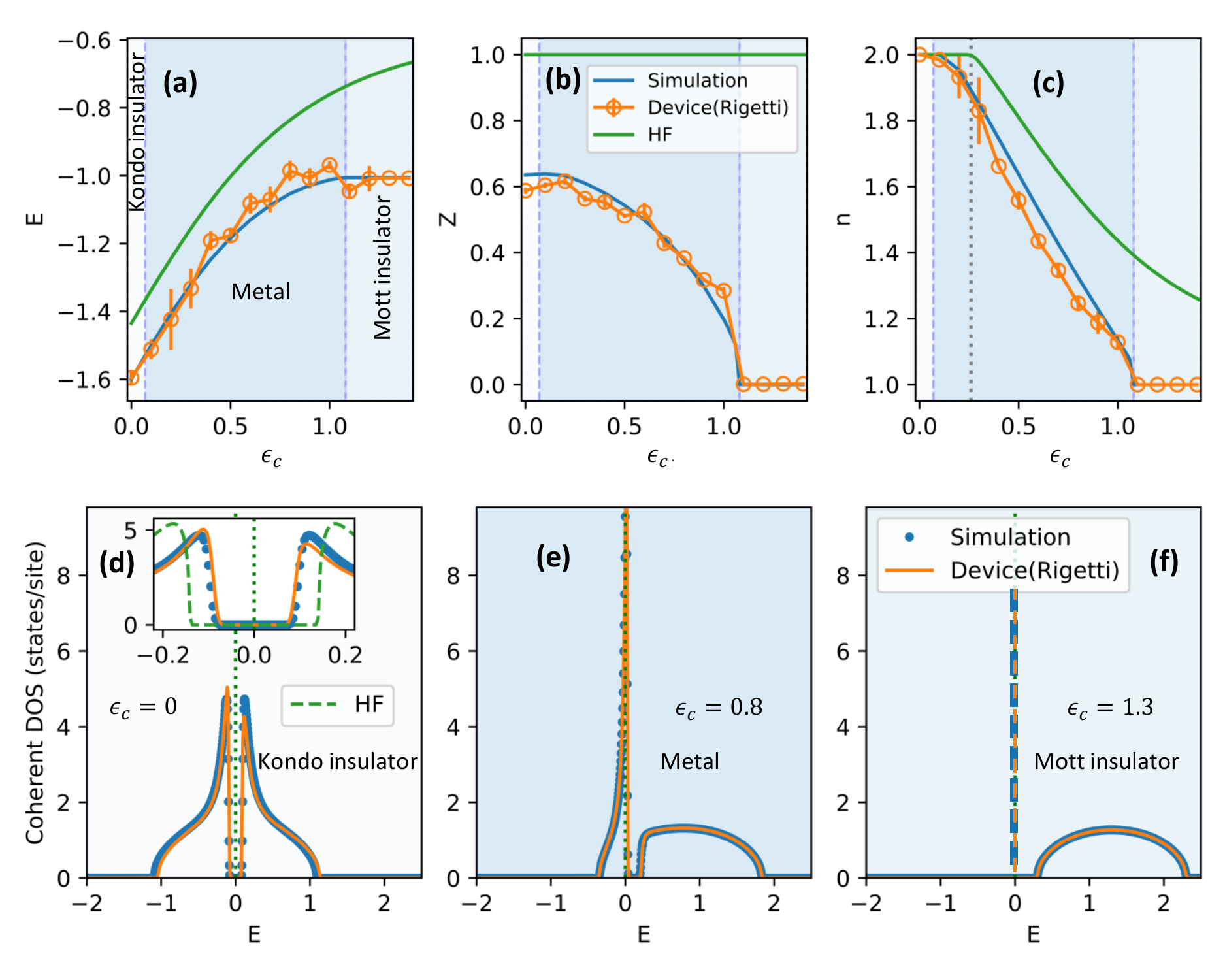}
	\caption{
	\textbf{GQCE results of quantum phases and phase transitions in the periodic Anderson model.} The Kondo insulator to metal and metal to Mott insulator electronic phase transitions are induced by raising the conduction band position $\epsilon_c$. Along the path, the variation of (a) total energy, (b) renormalization $Z$-factor, and (c) total electron filling \revision{per unit cell} $n$ are shown in the upper panels. We compare results from GQCE calculations using a VQE ansatz on (i) a noiseless statevector simulator (blue) and (ii) a real quantum device (Rigetti's Aspen-4) (yellow). We also show results of a purely classical Gutzwiller simulation using HF as the embedding Hamiltonian solver (green). The different phases (KI, M, MI) are presented in different color shadings with numerically exact phase boundaries taken from DMFT+NRG method~\cite{pam_logan2016}. The grey dotted line in panel (c) indicates the critical $\epsilon_c$ for Kondo insulator-metal transition described in HF theory. The lower panels show the coherent part of spectral density of states (DOS) of the Kondo insulator (d), metal (e) and Mott insulator (f) phases, obtained from GQCE calculations on the simulator and the real quantum device. The inset in panel (d) shows the DOS around the band gap with the HF results for comparison. \revision{The dashed vertical line in panel (f) indicates the coherent states at Fermi level with spectral weight diminishing to zero in Mott state.}
	}
	\label{figure4}
\end{figure*}

Figure~\ref{figure3} demonstrates the convergence of total energy, kinetic energy renormalization factor $Z\equiv \R^\dagger\R$, and maximal element of the error vector $\F$~\eqref{err} in our GQCE calculation on Aspen-4 as a function of iteration number. The iterative non-linear solver starts from the HF mean-field solution and reaches convergence after about 20 iteration steps. The remaining steps are used to estimate the error bars. The results using the real quantum device closely follow that of the noiseless simulations. The observed fluctuations stem from the device's noise. The maximal absolute value of the error vector elements in Fig.\ref{figure3}(c) levels near $0.01 (2\%)$, which coincides with the scale of the two-qubit CZ-gate fidelity of the device, which was about $95\%$. Because of the stochastic nature of quantum computing on real devices, hereafter we report results by mean values with estimated errors. The standard deviation is about $0.03 (2\%)$ for total energy and $0.01 (2\%)$ for $Z$-factor, estimated with the last 20 iterations in this calculation.

When the center of the conduction band is set to zero, $\epsilon_c=0$, as in Fig.~\ref{figure3}, the system is in the Kondo insulator phase. The local correlated $d$-orbital, which is also located at zero energy, hybridizes with the $c$-band and opens a Kondo gap. The $Z$-factor in Fig.~\ref{figure3}(b) shows appreciable amount of reduction from unity, manifesting the local on-site Coulomb interaction effect, which effectively reduces the hybridization energy.

Let us now consider the quantum phase diagram as we tune the position of the conduction band $\epsilon_c$. Even in this restricted parameter space, where all other parameters are held fixed, the PAM model goes through a series of quantum phase transitions from Kondo insulator to metal and from metal to Mott insulator. We compare our GQCE findings to the numerically exact phase boundaries at zero temperature that have been determined by DMFT calculations using NRG as the impurity solver~\cite{nrg, pam_logan2016}. To extract the phase boundaries, we calculate the change of total energy $E$, renormalization $Z$-factor and total electron filling $n$ as a function of $\epsilon_c$. Results are shown in the upper panels of Fig.\ref{figure4}, which also includes the numerically exact phase boundaries from DMFT+NRG for comparison.

As seen in Fig.~\ref{figure4}(a), the total energy from noiseless simulations monotonically increases with increasing $\epsilon_c$ and reaches a constant as the system crosses the metal-Mott insulator transition. The GQCE calculations on Aspen-4 follows closely the exact energy curve along the phase transformation path, yet with a sizable error bar that originates from the noise of the device. Gutzwiller theory offers an efficient treatment for the (orbital-selective) Mott insulating phase, which exploits the fact that Mott localized Gutzwiller quasi-particle bands are pinned at the chemical potential at integer filling~\cite{ga_uo2}. The embedding Hamiltonian in the Mott phase has a doubly degenerate ground state, which can be written as tensor product states $\ket{00}$ and $\ket{11}$ in the two-qubit parity basis. In practice, we choose one of the states to evaluate the energy and OPDM, followed by a symmetrization in the spin-sector to recover spin-symmetry.

We compare GQCE to HF calculations, where the embedding Hamiltonian solver is chosen to be at HF mean-field level (green curves in Fig.~\ref{figure4}). Within HF, the total energy is monotonically increasing and significantly larger than the GQCE result. Crucially, it bears no signature of the metal-insulator phase transitions. The important physical phenomenon that is not captured by HF theory is the suppression of energetically unfavorable doubly occupied sites in the Hilbert space of the correlated $d$-orbitals.

In Fig.~\ref{figure4}(b), we show the kinetic energy renormalization $Z$-factor~\cite{ga_qp}, which is a key physical concept captured by Gutzwiller theory. When the conduction band center rises above the zero chemical potential, the renormalization $Z$-factor drops gradually and vanishes at the metal to Mott-insulator transition. Remarkably, for the model parameters studied in this paper, GQCE predicts a metal-Mott insulator transition phase boundary that is in perfect agreement with the numerically exact value obtained from DMFT. The $Z$-factor obtained from GQCE calculations on the Aspen-4 quantum device closely follows the exact statevector simulation data. Within the HF approximation, the renormalization factor remains constant, $Z_{\text{HF}} = 1$, demonstrating that the metal-Mott insulator transition is beyond the description of HF theory.

Finally, in Fig.~\ref{figure4}(c), we show that the variation of the total electron filling per unit cell is an effective way to locate the phase boundaries. In the Kondo (Mott) insulator phases, the electron filling is equal to two (one), while it is in between the two values for the correlated metal phase. The electron filling obtained from GQCE calculations on Aspen-4 agrees well with the exact statevector simulations. \revision{Some underestimation in the middle range is present, manifesting the effect of noise in real devices.} The electron filling behavior can be used to locate the phase boundaries and the obtained critical parameter values are in agreement with the numerically exact ones. In contrast, the HF approach can only identify the transition from the Kondo insulator to the metal. As the correlation-induced renormalization of the hybridization is not captured within HF theory, the Kondo energy scale is overestimated and the Kondo insulator phase incorrectly persists up to larger values of $\epsilon_c$, (see dotted line in Fig.\ref{figure4}(c)).

The Gutzwiller method adopts a Jastrow-type variational wave function, which describes the ground state properties of a correlated model beyond an effective single-particle mean-field theory~\cite{gwf_review}. Although there is no efficient way currently available to evaluate the full Green's function within Gutzwiller approach, the coherent part of it can be straightforwardly calculated~\cite{ga_pu}. The resulting coherent spectral density of states (DOS), which includes coherent quasi-particle excitations, can be used to distinguish the different quantum phases in the model. The coherent DOS of the PAM model is shown in the lower panels of Fig.~\ref{figure4}(d-f), which correspond to Kondo insulator, correlated metal and Mott insulator phases. Data from GQCE calculations on Rigetti's Aspen-4 device are shown to be in excellent agreement with exact simulation results.

In the Kondo insulator phase (Fig.~\ref{figure4}(d)), the center correlated $d$-orbital hybridizes with the conduction band, resulting in a finite hybridization gap. The inset shows that the hybridization gap from GQCE calculations agrees well with the exact simulation result, and is significantly reduced compared with the HF mean-field value due to the correlation-induced renormalization of the hybridization strength $V \rightarrow \R V$. As the conduction band is lifted up to $\epsilon_c=0.8$, the system is situated in a metallic phase. The hybridization gap is still present but moves to higher energy, and the chemical potential is located at the sharp quasi-particle resonance peak. The total coherent spectral weight decreases in accordance with the smaller quasi-particle weight $Z$ as shown in Fig.~\ref{figure4}(b). At $\epsilon_c=1.3$, the coherent spectral weight completely vanishes as the $d$-orbital becomes Mott localized at half-filling. In the Mott phase, the incoherent lower and upper Hubbard bands, together with the conduction $c$-band, define the band gap size and distinguish between a Mott-Hubbard versus charge-transfer insulator phase. Although the GQCE calculations at this level cannot explicitly generate the Hubbard bands~\cite{gga_lanata2017}, the band gap size and characteristics can still be resolved by varying the chemical potential and monitoring the electron filling~\cite{ga_dmet}.

\revision{
\begin{figure}[ht]
	\centering
	\includegraphics[width=\linewidth]{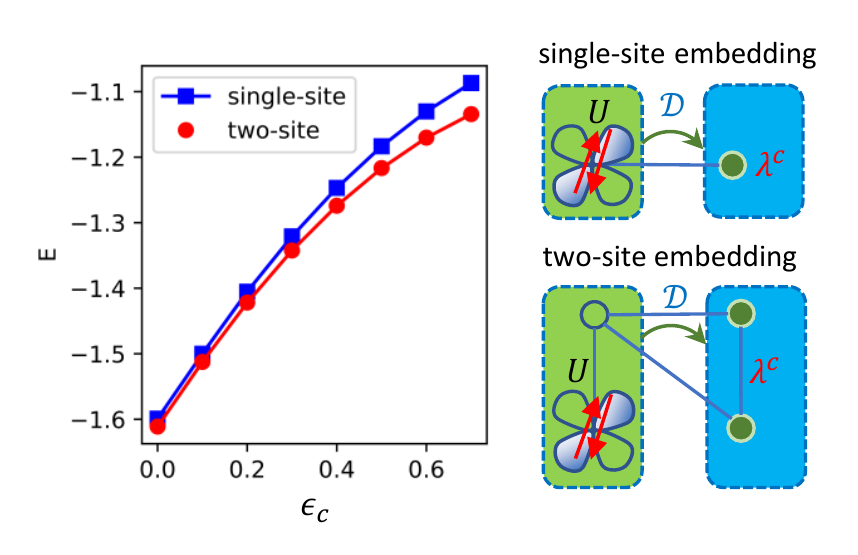}
	\caption{
	\revision{\textbf{GQCE calculations of PAM with spatially extended two-site Gutzwiller projector.} The left panel shows the total energy per unit cell $E$ of the PAM as a function of conduction band position $\epsilon_c$. Results are obtained from GQCE with single-site and two-site Gutzwiller projector. This includes a single correlated $d$-site and, for the two-site projector, also its nearest-neighbor uncorrelated $c$-site (see right panels). Symbols (lines) show GQCE results using VQE-UCCSD run on Qiskit state-vector simulator (exact diagonalization) as the embedding Hamiltonian solver~\cite{Qiskit}. The difference between ED and VQE results are smaller than the symbol size. The Hubbard interaction is set to $U=2$ in the calculation.}
	}
	\label{figure5}
\end{figure}

\section{Scaled up GQCE calculations}
As discussed in Sec.~\ref{variational_in_practice}, the GQCE approach is expected to maintain the variational nature in practice, and converge to the exact result by enlarging the Gutzwiller projector to include more nearby sites. For illustrations, we consider an extended Gutzwiller projector over the Hilbert space defined by both the correlated $d$-orbital site and the nearest uncorrelated $c$-orbital site in the PAM model such that spin-orbital dimension $N_\text{emb}$ increases from $4$ to $8$, as shown in Fig.~\ref{figure5}. The number of variational parameters in the UCCSD ansatz increases from effectively $1$ to $26$, and the number of two-qubit controlled-NOTs (CNOTs) in the VQE ansatz circuit increases from $2$ to $1096$. The GQCE total energy with the two-site Gutzwiller projector decreases due to the introduction of more variational degrees of freedom encoded in matrix $\Lambda_i$~\eqref{proj}. As an example, at $U=2$ and $\epsilon_c = 0.7$, the GQCE energy decreases by $4\%$ from $-1.0878$ to $-1.1343$ with VQE-UCCSD run on the statevector simulator, which agrees with the GQCE result with ED as the embedding Hamiltonian solver up to $5^\text{th}$ decimal place. Due to the fairly deep circuits with about $1000$ CNOTs, the GQCE calculation with the two-site Gutzwiller projector on the current noisy real device is still very demanding, which we will leave for future work.

For a general $f$-electron embedding Hamiltonian of spin-orbital dimension $28$ on the verge of classical computation limit, the VQE-UCCSD state preparation circuit requires about $1096\times (28/8)^5 \approx 5\times 10^5$ CNOTs and $\left( {7\choose 1}^2 + {7\choose 2}^2\right)\times 2 + {7\choose 1}^4 = 3381$ variational parameters, which poses a great challenge for the NISQ devices and classical optimizer. The hybrid quantum-classical optimization in VQE-UCCSD becomes more complicated with the observation of exponentially smaller probability of getting nonzero gradient at fixed precision with increasing number of qubits~\cite{mcclean2018barren}. Alternative approaches to VQE-UCCSD as the many-body embedding Hamiltonian ground state solver are available to be explored. The variational ansatz based on low-order Trotter approximation is simpler to implement and converges faster for model Hamiltonian calculations~\cite{wecker2015_trotterizedsp}. The adaptive VQE approaches have demonstrated to produce highly accurate ground state energies with much simpler variational circuits in quantum chemistry calculations~\cite{vqe_adaptive, vqe_qubit_adaptive, qcc_scott2018, iqcc_scott2020}. The quantum imaginary time evolution algorithm provides another axis to reach the ground state without the explicit complex high-dimensional optimizations~\cite{qite_chan20, VQITE, smqite, yao2020adaptive}. Furthermore, symmetries beyond the conservation of charge and spin, such as point group symmetries and approximate symmetries present in the embedding Hamiltonian, can also be utilized to taper off qubits or identify the relevant submanifold of the Hilbert space~\cite{bravyi2017tapering, setia2019reducing, FengVQE}.
}

\section{Conclusion}
To conclude, we have successfully implemented and benchmarked a novel hybrid quantum-classical simulation framework for interacting lattice models, which is based on the Gutzwiller variational embedding theory. In combination with density functional theory, this GQCE approach can describe ground state properties of correlated multi-orbital quantum materials. Using Rigetti's quantum cloud service, we have performed the first fully self-consistent hybrid quantum-classical calculation of an infinite correlated electron model on NISQ hardware. As a non-trivial benchmark study, we apply GQCE to the periodic Anderson model on the Bethe lattice using a single-site embedding scheme. We find excellent agreement between GQCE results obtained from Rigetti's Aspen-4 QPU and known numerically exact results.

The GQCE method lends itself well to NISQ technology as it maps the infinite lattice system to an effective, interacting impurity model, which is self-consistently coupled to a non-interacting fermionic bath. To obtain a self-consistent solution of a set of coupled eigenvalue equations, the method requires finding the ground state energy and single-particle density matrix of the impurity model, which can be done efficiently on QPUs. For the single \revision{and two-site} decoupling used here, we employ VQE with a unitary coupled cluster ansatz. \revision{We discuss the scaling of QPU resources with the size of the embedding cluster and conclude that larger impurity clusters may require using more efficient ans\"atze such as produced by adaptive VQE~\cite{vqe_adaptive} or using algorithm that bypass high-dimensional optimization such as the quantum imaginary time evolution method~\cite{qite_chan20,smqite}. }

Our work demonstrates the current capabilities of NISQ devices in simulating infinite lattice models of correlated materials. Even more importantly, exploiting the favorable linear scaling of the number of qubits with the size of the embedding Hamiltonian, we envision that VQE solutions of small $14$-site (28 spin-orbitals) Hubbard-type models will boost GQCE to the limit of what is currently possible on classical computers. This makes GQCE a promising framework for performing challenging computations of correlated materials in the near term, where NISQ devices may offer a practical quantum advantage.

\section*{Acknowledgements}
We acknowledge useful discussions with N. Lanat\`a, G. Kotliar and the Rigetti team, in particular with A. Brown, M. Reagor and M. Skilbeck. This work was supported by the U.S. Department of Energy (DOE), Office of Science, Basic Energy Sciences, Materials Science and Engineering Division. The research was performed at the Ames Laboratory, which is operated for the U.S. DOE by Iowa State University under Contract No. DE-AC02-07CH11358.

\section*{Data availability}
The GQCE calculations and data are available at figshare~\cite{gqce_pam}.

\section*{Code availability}
The GQCE code are available at figshare~\cite{pygqce}.

\bibliography{refabbrev, ref}

\end{document}


\newcommand{\E}{\mathcal{E}}
\newcommand{\G}{\mathcal{G}}
\newcommand{\Lag}{\mathcal{L}}
\newcommand{\M}{\mathcal{M}}
\newcommand{\N}{\mathcal{N}}
\newcommand{\U}{\mathcal{U}}
\newcommand{\R}{\mathcal{R}}
\newcommand{\F}{\mathcal{F}}
\newcommand{\V}{\mathcal{V}}
\newcommand{\C}{\mathcal{C}}
\newcommand{\I}{\mathcal{I}}
\newcommand{\s}{\sigma}
\newcommand{\sbar}{\bar{\sigma}}
\newcommand{\up}{\uparrow}
\newcommand{\dw}{\downarrow}
\newcommand{\h}{\hat{\mathcal{H}}}
\newcommand{\himp}{\hat{h}}
\newcommand{\g}{\mathcal{G}^{-1}_0}
\newcommand{\D}{\mathcal{D}}
\def\P{\mathcal{P}}
\newcommand{\A}{\mathcal{A}}
\newcommand{\K}{\textbf{k}}
\newcommand{\Q}{\textbf{q}}
\newcommand{\T}{\mathcal{T}}
\newcommand{\io}{i\omega_n}
\newcommand{\eps}{\varepsilon}
\newcommand{\+}{\dag}
\newcommand{\su}{\uparrow}
\newcommand{\giu}{\downarrow}

\newcommand{\ca}{\hat{c}^{\phantom{\dagger}}}
\newcommand{\cc}{\hat{c}^\dagger}
\newcommand{\fa}{\hat{f}^{\phantom{\dagger}}}
\newcommand{\fc}{\hat{f}^\dagger}

\newcommand{\aaa}{a^{\phantom{\dagger}}}
\newcommand{\aac}{a^\dagger}
\newcommand{\bba}{b^{\phantom{\dagger}}}
\newcommand{\bbc}{b^\dagger}
\newcommand{\da}{\hat{d}^{\phantom{\dagger}}}
\newcommand{\dc}{\hat{d}^\dagger}
\newcommand{\ha}{h^{\phantom{\dagger}}}
\newcommand{\hc}{h^\dagger}
\newcommand{\be}{\begin{equation}}
\newcommand{\ee}{\end{equation}}
\newcommand{\bea}{\begin{eqnarray}}
\newcommand{\eea}{\end{eqnarray}}
\newcommand{\ba}{\begin{eqnarray*}}
\newcommand{\ea}{\end{eqnarray*}}
\newcommand{\dagga}{{\phantom{\dagger}}}
\newcommand{\bR}{\mathbf{R}}
\newcommand{\bQ}{\mathbf{Q}}
\newcommand{\bq}{\mathbf{q}}
\newcommand{\bqp}{\mathbf{q'}}
\newcommand{\bk}{\mathbf{k}}
\newcommand{\bh}{\mathbf{h}}
\newcommand{\bkp}{\mathbf{k'}}
\newcommand{\bp}{\mathbf{p}}
\newcommand{\bL}{\mathbf{L}}
\newcommand{\bRp}{\mathbf{R'}}
\newcommand{\bx}{\mathbf{x}}
\newcommand{\by}{\mathbf{y}}
\newcommand{\bz}{\mathbf{z}}
\newcommand{\br}{\mathbf{r}}
\newcommand{\Ima}{{\Im m}}
\newcommand{\Rea}{{\Re e}}
\newcommand{\Pj}[2]{|#1\rangle\langle #2|}
\newcommand{\setof}[1]{\left\{#1\right\}}
\newcommand{\fract}[2]{\frac{\displaystyle #1}{\displaystyle #2}}
\newcommand{\Av}[2]{\langle #1|\,#2\,|#1\rangle}
\newcommand{\ov}[2]{\langle #1|\,#2\,\rangle}
\newcommand{\av}[1]{\langle #1 \rangle}
\newcommand{\Mel}[3]{\langle #1|#2\,|#3\rangle}
\newcommand{\Avs}[1]{\langle \,#1\,\rangle_0}

\newcommand{\Vb}{\bar{\mathcal{V}}}
\newcommand{\Vd}{\Delta\mathcal{V}}
\newcommand{\Rb}{\bar{R}}
\newcommand{\Rd}{\Delta R}

\renewcommand{\theequation}{S\arabic{equation}}
\renewcommand{\thefigure}{S\arabic{figure}}
\renewcommand{\bibnumfmt}[1]{[S#1]}
\renewcommand{\citenumfont}[1]{S#1}

\title{Supplemental Materials: Gutzwiller Hybrid Quantum Classical Simulation of Infinite Systems}
\author{Yongxin Yao}
\email{ykent@iastate.edu}
\affiliation{Ames Laboratory, U.S. Department of Energy, Ames, Iowa 50011, USA}
\affiliation{Department of Physics and Astronomy, Iowa State University, Ames, Iowa 50011, USA}

\author{Feng Zhang}
\affiliation{Ames Laboratory, U.S. Department of Energy, Ames, Iowa 50011, USA}



\author{Cai-Zhuang Wang}
\affiliation{Ames Laboratory, U.S. Department of Energy, Ames, Iowa 50011, USA}
\affiliation{Department of Physics and Astronomy, Iowa State University, Ames, Iowa 50011, USA}

\author{Kai-Ming Ho}
\affiliation{Ames Laboratory, U.S. Department of Energy, Ames, Iowa 50011, USA}
\affiliation{Department of Physics and Astronomy, Iowa State University, Ames, Iowa 50011, USA}

\author{Peter P. Orth}
\affiliation{Ames Laboratory, U.S. Department of Energy, Ames, Iowa 50011, USA}
\affiliation{Department of Physics and Astronomy, Iowa State University, Ames, Iowa 50011, USA}

\maketitle
Here we present main formalism of a generalized version of Gutzwiller variational embedding theory for multi-band systems developed in references~\cite{ga_pu,ga_uo2}, upon which we build the Gutzwiller hybrid quantum-classical simulation framework.

\section{Generic Hamiltonian}
Consider a generic multi-band interacting Hamiltonian for periodic systems of the following form
\be
\h = \sum_{\bk}\sum_{\mu\nu} \epsilon_{\bk\mu\nu}\, \cc_{\bk\mu}\ca_{\bk \nu} +\sum_{\bR i}\h_i^\text{loc}[\bR], \label{hamgen}
\ee
where $\bk$ is the crystal momentum, conjugate to the unit cell position vector $\bR$. $\mu,\nu$ are composite indices of lattice site, orbital and spin, with orbitals including both uncorrelated orbitals such as $s$ and $p$-orbitals, as well as correlated orbitals such as $d$ and $f$-orbitals. Likewise, the unit cell can refer to a primitive unit cell or a supercell. For convenience, we explicitly combine some sites and orbitals within the unit cell into groups labelled by index $i$, which are correlated due to the presence of Coulomb interactions. For example, a group $i$ can contain a fractionally occupied $d$-shell only, or plus the neighboring sites and orbitals to form a cluster. The Hamiltonian defined in the domain of $i$-group is denoted by
\be
\h_i^\text{loc}[\bR] = \sum_{AB} [\mathcal{H}_i^\text{loc}]_{AB} \Pj{A,\bR i}{B,\bR i}, 
\ee
which includes two-body Coulomb interactions and all the associated one-body terms such as crystal-field splitting and spin-orbit coupling. $A,B$ label the complete Fock state or occupation number basis set of the local Hilbert space defined by the spin-orbitals in the $i$-group, 
\be
\ket{A,\bR i} \equiv \prod_\alpha
    \left( \cc_{\bR i\alpha}\right)^{\Av{A}{\hat{n}_{\bR i \alpha}}}\ket{\emptyset},
\ee
with $\hat{n}$ the number operator and $\ket{\emptyset}$ the vacuum state. $\alpha$ runs through the site, orbital and spin in the $i$-group. The first component of Eq.~\ref{hamgen} contains the remaining part of the Hamiltonian, which are in quadratic form and mainly kinetic energy terms. The Hamiltonian can describe lattice model systems, such as Hubbard model, or real materials through mean-field electronic structure calculations such as density functional theory.

\section{Gutzwiller Variational Ansatz}
To solve for the ground state of the model beyond Hartree-Fock mean-field level, the GQC framework adopts the Gutzwiller variational wavefunction (GWF) ansatz,
\be
\ket{\Psi_\text{G}} = \prod_{\bR i} \hat{\P}_{\bR i} \ket{\Psi_0},
\ee
with 
\be
\hat{\P}_{\bR i} = \sum_{A,B}[\Lambda_i]_{A B}\Pj{A,\bR i}{B,\bR i}.
\ee
Variational parameters $\{ \Lambda_i \}$ have been introduced in the Gutzwiller projector $\hat{\P}_{\bR i}$ to optimize local correlated sectors of the noninteracting wavefunction $\ket{\Psi_0}$. A general square-matrix form of $\Lambda_i$ is enforced by the invariance of the total energy functional under unitary transformations in the linear subspace of orbitals in the $i$-group.

The expectation value of the Hamiltonian~\eqref{hamgen} with respect to GWF can not be evaluated analytically, except for some simple cases in certain limits~\cite{ga_1dvollhardt, ga_infdvollhardt}. Therefore, numerical results have to obtained through variational quantum Monte Carlo simulations without further approximation. To express the total energy functional in a semi-analytical form, which enables more efficient numerical calculations, we adopts the Gutzwiller approximation (GA)~\cite{ga, ga_bunemann1998, ga_fabrizio03, ga_pu}, which becomes exact for systems in infinite dimension or with infinite coordination number~\cite{ga_infdvollhardt}. GA is equivalent to rotationally invariant slave-boson method  with saddle-point approximation~\cite{sb_kotliar86, bunemann_gasb,ga_uo2}.

For nonlocal one-particle density operator or hopping operator $\cc_{\bR \mu} \ca_{\bR' \nu}$, which cannot be fully represented in the reduced local Hilbert space spanned by the spin-orbitals at a single group, the expectation value has a closed form as that of the noninteracting wavefunction subject to additional renormalization
\be
\Av{\Psi_\text{G}}{\cc_{\bR\mu} \ca_{\bR'\nu}} = \R_{a\mu}\R_{\nu b}^\dagger\Av{\Psi_0}{\fc_{\bR\mu}\fa_{\bR'\nu}}, \label{expval1}
\ee
with $\R$ the so-called Gutzwiller renormalization matrix. The quasiparticle $\fa$operator is introduced to distinguish it from the physical $\ca$operators defined in the Hamiltonian. The expectation value~\eqref{expval1} manifests the local electron correlation-induced renormalization effects on the kinetic energy of the systems.

For any local observable in $i$-group $\hat{O_i}[\{\cc_{i \alpha}\}, \{\ca_{i \alpha}\}]$, which is defined in a local correlated Hilbert space, the expectation value can be rigorously evaluated through the local reduced many-body density matrix, or equivalently the ground state wavefunction $\Phi_i$ of the Gutzwiller embedding Hamiltonian $\h^\text{emb}_i$ at site $i$
\be
\Av{\Psi_\text{G}}{\hat{O_i}} = \Av{\Phi_i}{\hat{O_i}},
\ee
with the embedding Hamiltonian of the following form
\bea
\h^\text{emb}_i [\D,\D^\dagger; \lambda^\text{c}] &=& \h^\text{loc}_i + \sum_{a \alpha}\left( [\D_i]_{a \alpha} \cc_{i \alpha} \fa_{i a} + h.c.\right)  \notag \\
    & & +\sum_{a b}[\lambda_i^\text{c}]_{a b} \fa_{i b} \fc_{i a}. \label{hemb}
\eea
The embedding Hamiltonian essentially describes a physical impurity $\h^\text{loc}_i$ coupled to a quadratic bath $\lambda^\text{c}$ of the same orbital dimension, with a coupling matrix $\D$.

The total energy of the system per unit cell can now be written as
\be
E = \Av{\Psi_0}{\hat{\T_\text{G}}[\R,\R^\dagger]} +\sum_{i} \Av{\Phi_i}{\h^{\text{loc}}_i},
    \label{etot}
\ee
with
\be
\hat{\T_\text{G}}[\R,\R^\dagger] = \frac{1}{N_\bk}\sum_{\bk}\sum_{\mu\nu}\sum_{a b}
    \epsilon_{\bk\mu\nu} \R_{a\mu}\R_{\nu b}^\dagger
    \fc_{\bk a}\fa_{\bk b}. 
\ee
It consists of contribution from the expectation value 
of an effective quasi-particle Hamiltonian 
and that of the quantum embedding Hamiltonians. 

\section{Lagrange equations}
The total energy in Eq.~\ref{etot} is to be minimized in the parameter space defined the noninteracting wavefunction $\Psi_0$ and local Hilbert space of each nonequivalent embedding Hamiltonian, subject to the Gutzwiller constraints~\cite{ga_bunemann1998, ga_fabrizio07}
\bea
\ov{\Phi_i}{\Phi_i} &=& 1, \\
\Av{\Phi_i}{\fa_{i b}\fc_{i a}} &=& \Av{\Psi_0}{\fc_{i a}\fa_{i b}}. 
\eea
The constrained minimization can be conveniently formulated 
with the following Lagrange function
\be
\begin{split}
&\Lag[\Psi_0,E^\text{p}; \Phi,E^\text{c}; \R,\R^\dagger,\lambda; \D,\D^\dagger,\lambda^\text{c}; \Delta_\text{p}] 
    = \Av{\Psi_0}{\hat{\T_\text{G}}[\R,\R^\dagger]} + E^\text{p}(1-\ov{\Psi_0}{\Psi_0}) \\
    &\;\;+\sum_i \left[ \Av{\Phi_i}{\h^\text{emb}_i [\D,\D^\dagger; \lambda^\text{c}]}
    + E_i^\text{c} (1-\ov{\Phi_i}{\Phi_i})\right] \\
    &\;\;-\sum_i \left[ 
    \sum_{a b} \left( [\lambda_i]_{a b} + [\lambda_i^{c}]_{a b} \right)
    [\Delta_{\text{p} i}]_{a b}
    +\sum_{c a \alpha}\left( [\D_i]_{a\alpha}[\R_i]_{c\alpha}
    \left[\Delta_{\text{p} i}(1-\Delta_{\text{p} i}) \right]^{\frac{1}{2}}_{c a} +c.c. \right)
    \right].
\end{split}
\ee
The minimization of the Lagrange function with respect to 
all of the independent variables leads to 
the following Lagrange equations
\bea
&&\h_\text{G}^\text{qp}[\R,\R^\dagger; \lambda]\ket{\Psi_0} = E^\text{p} \ket{\Psi_0}, \\
&&\Av{\Psi_0}{\fc_{i a}\fa_{i b}} = [\Delta_{\text{p} i}]_{a b}, \\
&&\Av{\Psi_0}{\frac{\partial \hat{\T_\text{G}}[\R,\R^\dagger]}{\partial \R_{i a\alpha}}} 
    = \sum_c [\D_i]_{c\alpha} \left[\Delta_{\text{p} i}(1-\Delta_{\text{p} i})\right]
    ^\frac{1}{2}_{a c}, \\
&&\sum_{c b\alpha}\frac{\partial}{\partial d_{i s}^\text{p}}
    \left[\Delta_{\text{p} i}(1-\Delta_{\text{p} i})\right]^\frac{1}{2}
    _{c b}[\D_i]_{b\alpha}[\R_i]_{c\alpha}
    + c.c. + [l+l^\text{c}]_{i s} = 0, \\
&&\h^\text{emb}_i [\D,\D^\dagger; \lambda^\text{c}]\ket{\Phi_i} = E_i^\text{c} \ket{\Phi_i}, \\
&&\left[ \F_i^{(1)}\right]_{a\alpha} \equiv
    \sum_c\left[\Delta_{\text{p} i}(1-\Delta_{\text{p} i})\right]^{-\frac{1}{2}}_{c a}  \Av{\Phi_i}{\cc_{i\alpha}\fa_{i c}} - [\R_{i}]_{a\alpha} = 0, \\
&&\left[ \F_i^{(2)}\right]_{ab} \equiv \Av{\Phi_i}{\fa_{i b}\fc_{i a}} 
    - [\Delta_{\text{p}i}]_{ab} = 0,
\eea
with 
\be
\h_\text{G}^\text{qp}[\R,\R^\dagger; \lambda] \equiv \hat{\T_\text{G}}[\R,\R^\dagger] 
    + \sum_{i a b}[\lambda_i]_{ab}\fc_{i a}\fa_{i b}.
\ee
In the derivation, we have expanded the local one-particle matrix $\lambda$,
$\lambda^\text{c}$, and $\Delta_\text{p}$ in a complete set of local symmetry-adapted 
orthonormal Hermitian matrix basis $\{h_{is}\}$ as follows
\bea
&&\lambda_i = \sum_s l_{is} h_{is}, \\
&&\lambda_i^\text{c} = \sum_s l_{is}^\text{c} h_{is}, \\
&&\Delta_{\text{p}i} = \sum_s d_{is}^\text{p} [h^T]_{is}.
\eea

\section{Parity representation of the molecule Hamiltonian}
The Gutzwiller embedding Hamiltonian~\eqref{hemb} can be rewritten as, apart from a constant,
\be
\h^\text{emb} = \sum_{p q\s}t_{p q}\hat{\phi}_{p\s}^\dag \hat{\phi}_{q\s} + \frac{1}{2}\sum_{p q r s}\sum_{\s \s'} U_{p q r s}\hat{\phi}_{p\s}^\dag \hat{\phi}_{r\s'}^\dag \hat{\phi}_{s\s'} \hat{\phi}_{q\s}, \label{hemb1}
\ee
where $p, q, r, s$ labels the correlated $d$ and bath $f$-orbitals. For the periodic Anderson model (PAM) presented in the main text,  the $2\times 2$ matrix $t$ has the form
\be
t = 
\begin{bmatrix}
\epsilon_{d} & \D\\
\D & -\lambda^{c}
\end{bmatrix}.
\ee
The Coulomb tensor $U$ is quite sparse with nonzero elements only between correlated $d$-orbitals, which is $U_{1111}$ in the PAM model. To facilitate UCCSD-VQE calculations, a self-consistent Hartree-Fock (HF) calculation is performed for the embedding Hamiltonian. The orthonormal molecular orbitals can be generally express as $\chi_p = \sum_{p'}V_{pp'}\phi_{p'}$. For PAM, The $2\times 2$ unitary matrix $V$ has the following form:
\be
V = 
\begin{bmatrix}
\cos(x) & \sin(x)\\
-\sin(x) & \cos(x)
\end{bmatrix},
\ee
where the angle $x$ is determined by the HF solution. The embedding Hamiltonian~\eqref{hemb1} is then transformed from $\{d,f\}$-basis to $\chi$-basis:
\be
\h^\text{emb} = \sum_{p q \s}\sum_{p' q'}V_{pp'}t_{p' q'}V_{q' q}^\dag \hat{\chi}_{p\s}^\dag \hat{\chi}_{q\s} + \frac{1}{2}\sum_{p q r s}\sum_{\s \s'} \sum_{p' q' r' s'}V_{qq'}V_{pp'}U_{p' q' r' s'}V_{r' r}^\dag V_{s' s}^\dag \hat{\chi}_{p\s}^\dag \hat{\chi}_{r\s'}^\dag \hat{\chi}_{s\s'} \hat{\chi}_{q\s}. \label{hemb2}
\ee
Note that the sparsity of $U$-matrix is generally removed in the HF orbital representation. Standard procedures can be followed to transform the embedding Hamiltonian in Eq.~\eqref{hemb2} from fermionic representation to qubit representation using Jordan-Wigner encoding~\cite{map_jw}, parity encoding or  Bravyi-Kitaev encoding~\cite{map_bk, seeley2012bravyi}. 

For PAM, the embedding Hamiltonian is reduced to the following form with parity encoding and two-qubits tapered off due to spin and particle number conservation:
\be
\h^\text{emb} = g_0 \mathds{1} + g_1 (Z_0-Z_1) + g_2 (X_0+X_1) + g_3 Z_0 Z_1 + g_4 (X_0Z_1 - Z_0 X_1) + g_5 X_0 X_1\,.
\ee
As a numerical example, at $U=2$ and $\epsilon_c = 0.5$, the GQCE solution leads to an embedding Hamiltonian with
\be
t = 
\begin{bmatrix}
-1 & -0.265\\
-0.265 & 0.095
\end{bmatrix}.
\ee
The Hartree-Fock self-consistent solution yields two molecular orbitals
\be
V = 
\begin{bmatrix}
0.729 & 0.685 \\
-0.685 & 0.729
\end{bmatrix}.
\ee
With a basis transformation to molecular orbital representation, $t$ changes to
\be
V t V^\dag = 
\begin{bmatrix}
-0.751 & 0.530\\
0.530 & -0.154
\end{bmatrix}.
\ee
and the Coulomb integral $U$ changes from a sparse tensor $[[[[2.0, 0.0], [0.0, 0.0]],$ $ [[0.0, 0.0], [0.0, 0.0]]], $ $ [[[0.0, 0.0], [0.0, 0.0]], $ $ [[0.0, 0.0], [0.0, 0.0]]]]$ in atomic orbital representation, to a dense tensor $[[[[0.564, -0.53], [-0.53, 0.498]],$ $ [[-0.53, 0.498], $ $ [0.498, -0.468]]], $ $ [[[-0.53, 0.498], $ $ [0.498, -0.468]], $ $ [[0.498, -0.468], [-0.468, 0.44]]]]$ in molecular orbital representation. In parity encoding the embedding Hamiltonian becomes
\be
\h^\text{emb} = -0.405 \mathds{1} -0.267 (Z_0 - Z_1)  + 0.031 (X_0 + X_1) - 0.002 Z_0 Z_1 + 0.031 (X_0Z_1 -Z_0 X_1) + 0.498 X_0 X_1\,.
\ee

\section{Hybrid quantum-classical computing}
The Gutzwiller variational approach has been shown to be 
highly effective to describe ground state properties 
of correlated electron systems
\cite{ga_ce,ga_fe,ga_feas_bunemann,ga_pu,ga_uo2,ga_smb6,ga_tmo}.
For correlated multi-orbital systems, 
the ground state solution of the Gutzwiller embedding Hamiltonian 
becomes the bottleneck for computational load in classical computers.
However, quantum computers offer a potentially exponential speedup 
in solving ground state and dynamics of quantum systems, 
taking advantage of the superposition and entanglement in the qubit space.
With the fast development of the near-term noisy quantum (NISQ) computing technology,
various hybrid quantum eigensolvers with feasible shallow quantum circuits
have been proposed.
Applications to simple molecules have been demonstrated on real NISQ devices
~\cite{vqe, hardware_efficient_vqe, vqe_pea_h2, vqe_NaH}.
The Gutzwiller quantum embedding Hamiltonian in Eq.\ref{hemb}, 
in a form similar to finite molecule 
but with simplified two-body Coulomb interaction, 
can naturally be solved using NISQ devices. 

\begin{figure}[ht]
	\centering
	\includegraphics[width=3.5in]{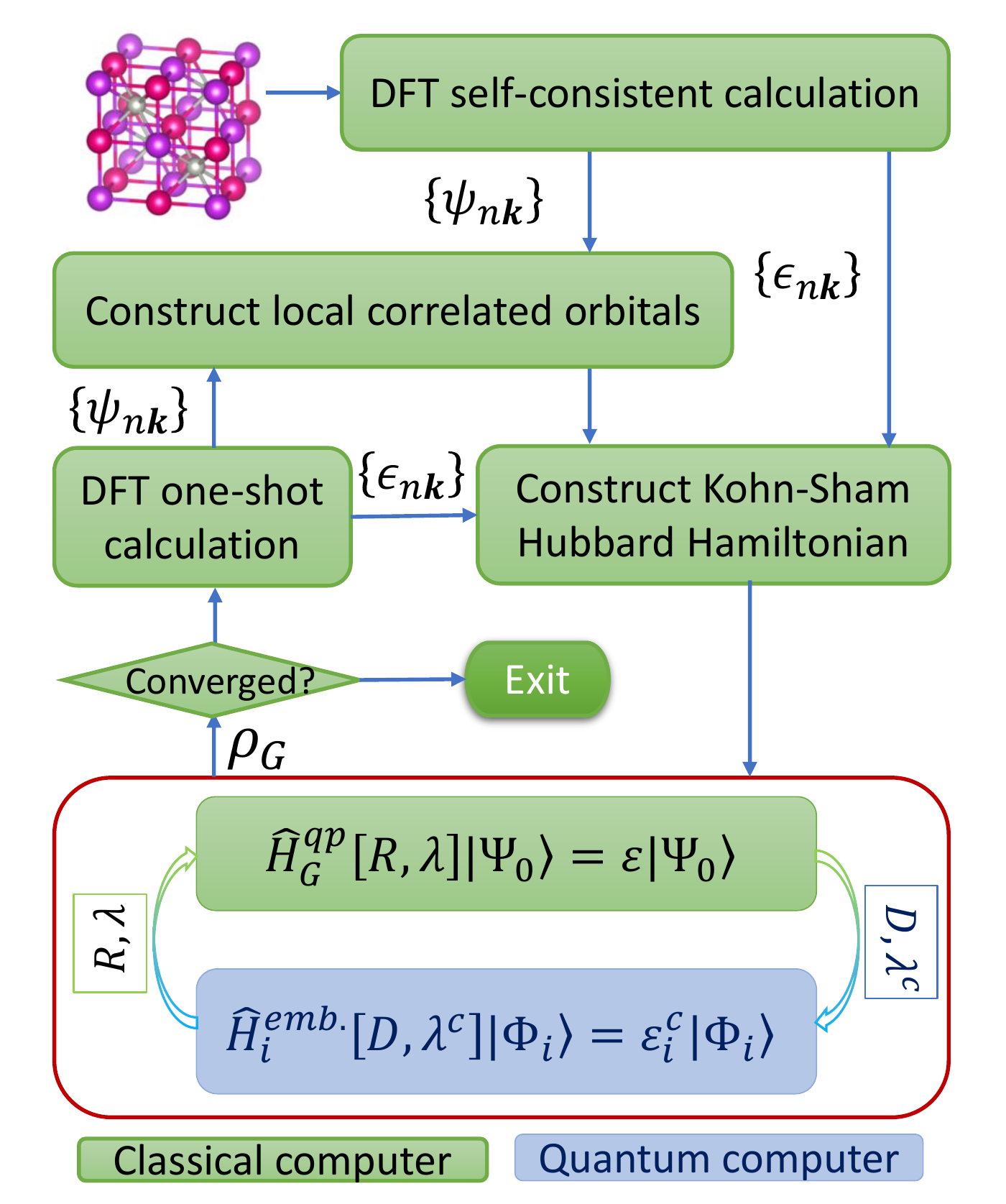}
	\caption{
	Flowchart of hybrid quantum-classical simulations of real materials 
	within DFT+Gutzwiller approach.
	}
	\label{figure1}
\end{figure}

Gutzwiller theory can be combined with first-principles electronic structure 
method such as DFT, 
and readily simulate real correlated materials. 
Figure \ref{figure1} illustrates the algorithmic flow 
of the GQC calculations of real materials, 
in combination with DFT.
Starting from the crystal structure of real material,
self-consistent DFT calculation produces the band energies $\{\epsilon_{n\bk}\}$
and band wave functions $\{\psi_{n\bk} \}$,
which defines the quadratic part of the model.
The generic Hubbard model is obtained by adding explicitly 
the local onsite Coulomb interactions for the correlated atomic shells,
which is subsequently solved following the above Gutzwiller ansatz. 
Quantum computers are used to solve the ground state of 
the Gutzwiller embedding Hamiltonian 
for great potential scalability and efficiency. 
The solution of the Gutzwiller Lagrange equations leads to 
renormalized electron density,
which is fed back to DFT for further iteration until convergence.
Such Gutzwiller hybrid quantum-classical simulation framework will 
extend the impact of NISQ technologies to infinite systems, 
addressing condensed matter physics and material science problems.

\bibliography{refabbrev, ref}